# Plasma-Cascade Instability- theory, simulations and experiment

Vladimir N. Litvinenko*[1,2], Gang Wang[2,1], Yichao Jing[2,1], Dmitry Kayran[2,1], Jun Ma[2], Irina Petrushina[1], Igor Pinayev[2] and Kai Shih[1]

[1] Department of Physics and Astronomy, Stony Brook University, Stony Brook, NY
[2] Collider-Accelerator Department, Brookhaven National Laboratory, Upton, NY

*Abstract.* In this letter we describe a new micro-bunching instability occurring in charged particle beams propagating along a straight trajectory: based on the dynamics we named it a Plasma-Cascade Instability. Such instability can strongly intensify longitudinal micro-bunching originating from the beam's shot noise, and even saturate it. Resulting random density and energy microstructures can drastically reduce beam's quality. Conversely, such instability can drive novel high-power sources of broadband radiation or can be used as a broadband amplifier. We discovered this phenomenon in a search for such amplifier in Coherent electron Cooling scheme [Phys. Rev. Lett. 102, 114801 (2009)] without separation of electron and hadron beams. In this letter we present our analytical and numerical studies of this new phenomenon as well as the results of its experimental demonstration.

High brightness intense charged particle beams play critical role in the exploration of modern science frontiers [1]. Such beams are central for high luminosity hadron colliders [2-5] as well as for X-ray femtosecond free-electron-lasers (FEL) [6-20]. In the future, such beams could be central for cooling hadron beams in high-luminosity colliders [21-23], X-ray FEL oscillators [24-27], and plasma-wake-field accelerators with TV/m accelerating gradients [28-36].

Preservation of the beam quality during generation, acceleration, transportation and compression is important for attaining the desirable properties of the beam. Dynamics of high intensity beams is driven by both external factors—such as focusing and accelerating fields—and self-induced (collective) effects: space charge [37-58], wakefields from the surrounding environment and radiation of the beam [59-65]. While external factors are designed to preserve beam quality, the collective effects can produce an instability [59] severely degrading beam emittance(s)[1], momentum spread and creating filamentation of the beam. On the other hand, such instabilities can be deliberately built-in to attain specific results. The most known application is the FEL instability used for generating coherent radiation from THz to X-rays [66-70]. Less known applications are Coherent electron Cooling (CeC) of hadron beams [71-75] or generation of broad-band high power radiation [76-78].

The Plasma-Cascade micro-bunching Instability (PCI) occurs in a beam propagating along a straight line. It is driven by variation of the transverse beam size(s) [75]. Conventional micro-bunching instability for beams travelling along a curved trajectory[2] is a well-known and in-depth studied both theoretically and experimentally [79-97]. Space-charge-driven parametric transverse instabilities are also well known (see review [98] and references therein). But none of them include the PCI—a micro-bunching <u>longitudinal instability driven by modulations of the transverse beam</u>

---

[1] Beam emittance is the phase space area (for 1D case) or the phase space hyper-volume (for 2D and 3D case) occupied by the beam's particles.

[2] For example, in a magnetic chicane or in an arc of an accelerator.

size. We start from a qualitative description of the PCI, which will be followed by theory, 3D simulations and experimental observations of this phenomenon.

Fig. 1 depicts a periodic focusing structure where the charged particle beam undergoes periodic variations of its transverse size. It is known that small density perturbations $\tilde{n}(\vec{r}), |\tilde{n}| \ll n_o$ in a cold, infinite and homogeneous charged beam will undergo oscillations with plasma frequency, $\omega_p = c\sqrt{4\pi n_o r_c}$ [99], where $n_o$ is the particles density (in beam's co-moving frame), $c$ is the speed of light and $r_c = e^2/mc^2$ is particle's classical radius. Beam propagating with velocity $v_o$ through the lattice (with period $2l$) would experience density modulation in the co-moving frame with period of $T = 2l/\gamma_o v_o$:

$$f_\perp = n_o(t) = \frac{I_o}{e\beta_o \gamma_o c} \frac{1}{\pi a^2(\gamma_o \beta_o ct)}. \tag{1}$$

where $I_o$ is the beam current and $\gamma_o = (1-\beta_o^2)^{-1/2}, \beta_o = v_o/c$ is the beam's relativistic factor.

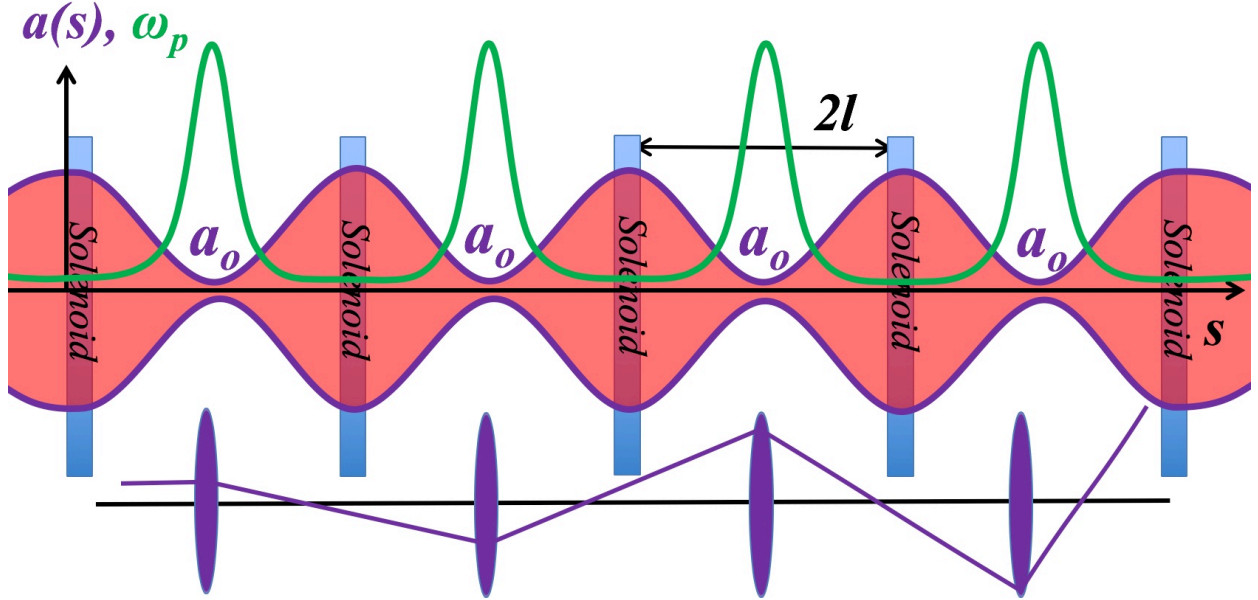

Figure 1. A sketch of four focusing cells with periodic modulations of beam envelope, *a(s)*, and the plasma frequency, *ω$_p$*. Beam envelope has waists, *a$_o$*, in the middle of each cell where plasma frequency peaks. Scales are attuned for illustration purpose. The bottom sketch illustrates an unstable ray trajectory in a system of periodic focusing lenses—an analog of unstable longitudinal oscillations. The waists of the beam serve as "short focusing elements" for the longitudinal plasma oscillations.

It is well known in classical oscillator theory [100] that modulation of oscillator frequency with a period close to a half of oscillation period would result in exponential growth of oscillation amplitude: the phenomena known as parametric resonance. Both the width and growth rate of this parametric instability can be enormous (see Fig. 2) when the amplitude of frequency modulation is large. The extreme case of δ-function-like modulation is well known: periodic focusing lenses with focal length shorter than a quarter of the separating distances will make rays unstable and the entire half-space $F < l/2$ is occupied by this parametric resonance.

These two examples illustrated that modulation of the transverse size, either periodic or aperiodic, could lead to an exponential growth of longitudinal density modulation: e.g. a solution of s-dependent Hill's equation $x'' + K(s)x = 0$ can be unstable for $K(s) \geq 0$.

*Analytical studies.* To switch from qualitative to quantitative description of PCI we need to identify a relevant analytically tractable problem. In mathematical terms, we need to separate transverse and longitudinal degrees of freedom: $f = f_\perp \cdot f_\parallel$. For compactness, we will consider only case of periodic axially-symmetric systems with round beams: case of elliptical beam in aperiodic lattice will have the same underlaying PCI physics while requiring a lengthy write-up. Following [75] we will consider a long bunch $\sigma_s \gg a/\gamma_o$ with transverse Kapchinsky-Vladimirsky (KV) distribution [101] providing uniform density inside the beam envelope, $r \leq a(s)$, and zero density outside it [102]. The beam envelope is described by a second-order nonlinear differential equation [102]:

$$\frac{d^2 a}{ds^2} + K(s)a - \frac{2}{\beta_o^3 \gamma_o^3} \frac{I_o}{I_A} \frac{1}{a} - \frac{\varepsilon^2}{a^3} = 0; \quad K(s) = \left(\frac{eB_{sol}(s)}{2p_o c}\right)^2, \quad (2)$$

where $\varepsilon$ is the envelope emittance[3], $p_o = \gamma_o \beta_o mc$ is particles' momentum, $K$ is external focusing provided by solenoid's field $B_{sol}$, and $I_A = mc^3/e \approx 17\ kA$ is the Alfven current. With a few exceptions [75] eq. (2) has to be solved numerically. Its solution defines the uniform time-dependent transverse distribution function $f_\perp$ given by eq. (1).

A weak perturbation of longitudinal distribution function $f_\parallel = f_o(v) + \tilde{f}(z,v,t), |\tilde{f}| < f_o$ (v is z-component of velocity in the co-moving frame) can be expressed through its Fourier components: $\tilde{f} = \int \tilde{f}_k(v,t)e^{ikz} dk$. Evolution of a high frequency component, $\tilde{f}_k$ ($ka \gg 1$) [75,103], is described by a linearized set of Vlasov-Poisson equations [104]:

$$\frac{\partial}{\partial t}\tilde{f} + ikv\tilde{f}_k + \frac{eE_z}{m} \cdot \frac{\partial f_o}{\partial v} = 0; \quad ikE_z = 4\pi e n_o(t) \cdot \int_{-\infty}^{\infty} \tilde{f}_k(v,t) v\, dv. \quad (3)$$

Following the technique developed in [105-106] and assuming κ-1 longitudinal velocity distribution $f_o(v) = \sigma_v / \pi(\sigma_v^2 + v^2)$ one can reduce (3) to an oscillator equation:

---

[3] It's worth mentioning that in the KV distribution:

$$f_\perp(x,x',y,y') = N \cdot \delta\left(\frac{x^2 + y^2}{a^2} + \frac{(ax' - a'x)^2 + (ay' - a'y)^2}{\varepsilon^2} - 2\right); a' = \frac{da_u}{ds},$$

the emittance $\varepsilon$ describes particles at the edge of the beam and is 4-fold larger than RMS beam emittance.

$$\frac{d^2 \tilde{g}_k}{dt^2} + \omega_p^2(t) \tilde{g}_k = 0; \; \tilde{g}_k = \exp(|k\sigma_v t|) \int_{-\infty}^{\infty} \tilde{f}_k(v,t) dv; \qquad (4)$$

for density perturbation corrected by Landau damping $\exp(|k\sigma_v t|)$ term [107-108]. Solution of Hill's eq. (4) can be expressed using symplectic ($det\,\mathbf{M} = 1$) transport matrix:

$$X(t) = \mathbf{M}(0|t) \cdot X_o; \; \mathbf{M}(0|t) = \exp\left(\int_0^t \mathbf{D}(\tau) d\tau\right); \; \mathbf{D}(t) = \begin{bmatrix} 0 & 1 \\ -\omega_p^2(t) & 0 \end{bmatrix}. \qquad (5)$$

In a periodic system comprised of $m$ bilaterally-symmetric cells[4], such as in Fig. 1, evolution is determined by eigen values of the cell transport matrix $\mathbf{M}_c = \mathbf{M}(0|T)$, $\mathbf{M}(0|mT) = \mathbf{M}_c^m$ [109]:

$$\mathbf{M}_c = \begin{bmatrix} m_{11} & m_{12} \\ m_{21} & m_{11} \end{bmatrix}; \; \lambda_1 = \lambda_2^{-1} = m_{11} - \sqrt{m_{11}^2 - 1};$$

$$\tilde{g}_k(mT) = \frac{\lambda_1^m + \lambda_1^{-m}}{2} \tilde{g}_k(0) - \frac{m_{12}}{\sqrt{m_{11}^2 - 1}} \frac{\lambda_1^m - \lambda_1^{-m}}{2} \dot{\tilde{g}}_k(0), \qquad (6)$$

with $|m_{11}| > 1$ indicating exponential growth, e.g. instability. Additional utility of periodic (cell) structure is that equations (2) and (4) can be reduced to a dimensionless form:

$$\frac{d^2 \hat{a}}{d\hat{s}^2} - k_{sc}^2 \hat{a}^{-1} - k_\beta^2 \hat{a}^{-3} = 0; \; \frac{d^2}{d\hat{s}^2} \tilde{n}_k + 2 \frac{k_{sc}^2}{\hat{a}(\hat{s})^2} \cdot \tilde{n}_k = 0; \; \hat{a} = \frac{a}{a_o}; \; \hat{s} = \frac{s}{l}, \qquad (7)$$

with variables $\hat{a} \geq 1$; $\hat{s} \in \{-1,1\}$ and the system dynamics defined by two parameters representing space charge and emittance effects:

$$k_{sc} = \sqrt{\frac{2}{\beta_o^3 \gamma_o^3} \frac{I_o}{I_A} \frac{l^2}{a_o^2}}; \; k_\beta = \frac{\varepsilon l}{a_o^2}.$$

Fig. 2 shows the growth rate per cell in such system evaluated by a semi-analytical code in Mathematica [110] using 4-th order symplectic integrators[5] [111].

---

[4] Bilateral symmetry of the cell provides for equality of the diagonal elements of $\mathbf{M}_c$: $m_{11} = m_{22}$.

[5] Equations (7) are canonical with the following Hamiltonians: $\hat{h} = \frac{\hat{a}'^2}{2} - k_{sc}^2 \ln \hat{a} + \frac{k_\beta^2}{2\hat{a}^2} = \frac{k_\beta^2}{2} = inv$

; $\tilde{h}(s) = \frac{\tilde{n}_k'^2}{2} + \frac{k_\beta^2}{\hat{a}^2(s)} \tilde{n}_k^2$.

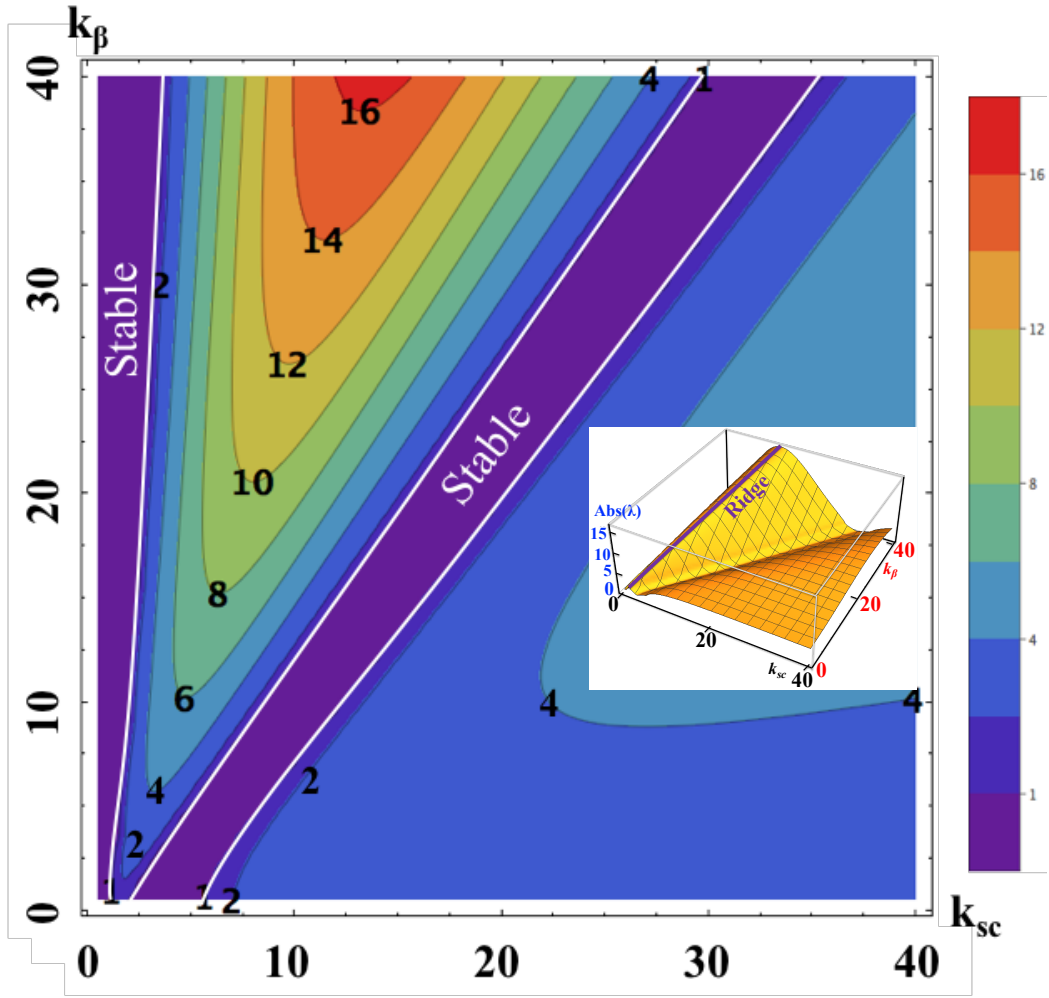

Figure 2. Contour plots of $\lambda = \max(|\mathrm{Re}\,\lambda_1|, |\mathrm{Re}\,\lambda_2|)$, absolute value of the growth rate per cell. Purple area highlighted by white lines indicates the areas of the stable oscillations $|\lambda_{1,2}| = 1$. Density modulation grows exponentially outside these areas. The 3D form of this graph in the insert shows clearly identifiable ridge along the $k_\beta = 3 \cdot (k_{sc} - 1.2)$ line, where growth rates peak.

The growth rates peak along the ridge $k_\beta = 3 \cdot (k_{sc} - 1.2)$ and can be estimated there as $\lambda \propto 1.25 k_{sc} \approx 1.5 + 0.413 k_\beta$. Space charge alone does not generate growth exceeding $\lambda \sim 4$.

While $\lambda$ does not contain any dependence on the modulation wavenumber $k$ [6], the PCI growth is inhibited both at high and low $k$. At high $k$ the growth is limited by Landau damping (7) to

$$k_{max} = \frac{\ln \lambda}{T \sigma_v}; \quad \omega_{max} = \frac{v_o}{2l} \cdot \frac{\gamma_o^3}{\sigma_\gamma} \ln \lambda,$$

---

[6] Corresponding frequency of modulation in the lab-frame is $\omega_m = \gamma_o v_o k$.

where $\sigma_\gamma / \gamma_o$ is the relative energy spread in the beam: $\sigma_\gamma / \gamma_o = \sigma_v / c$. The energy spread should take into account angular spread in the beam, which changes longitudinal velocity of the particle. Analytical studies of modulation at low $k$ are rather elaborate (see [75]) and here we provide an approximate scaling of space charge coefficient at $k\langle a(s)\rangle < 1$: $k_{sc} \to k^*_{sc} = k_{sc} \cdot (k\langle a(s)\rangle)$. The PCI extincts at low $k$ when the "working point" $k^*_{sc}, k_\beta$ reaches stable region.

*Numerical studies.* While both qualitative and semi-analytical studies reveal the nature of the PCI, modern particle-in-cell codes allow 3D investigations of PCI without any predetermined assumptions. We used code SPACE [112-113] for accurate simulation of PCI in electron beam with constant beam energy propagating along the straight section with focusing solenoids. Using this code, we confirmed that indeed the PCI could occur in periodic and aperiodic beamlines for a wide range of parameters, modulation frequencies and beam energies. Three sets of parameters listed in Table 1 were used in our studies. Fig. 3 shows simulation results for two test cases of the periodic lattices with PCI at 25 THz and 1 PHz (1,000 THz). Case of the aperiodic lattice is related to our experiment and is described in the next section, and illustrated in Figs. 5 and 6.

Table 1. Electron beam parameters for PCI simulations and tests

| Name | Exp | Case 1 | Case 2 |
|---|---|---|---|
| Lattice | Aperiodic LEBT | Periodic 4 cells | Periodic 4 cells |
| $\gamma$ | 3.443 | 28.5 | 275 |
| **E, MeV** | **1.76** | **14.56** | **140.5** |
| $l$, m | 1.5 - 3 | 1 | 10 |
| $a_0$, mm | 0.3, min | 0.2 | 0.1 |
| $I_0$, A | 1.75 | 100 | 250 |
| $\varepsilon_{norm}$, m | 1 10$^{-6}$ | 8 10$^{-6}$ | 4 10$^{-4}$ |
| $k_{sc}$ | Varies | 3.56 | 3.76 |
| $k_\beta$ | Varies | 7.02 | 14.55 |
| $\lambda_1$, per cell | N/A | -4.06 | -5.10 |
| Energy spread | 1 10$^{-4}$ | 1 10$^{-4}$ | 1 10$^{-4}$ |
| **Frequency, THz** | **0.4** | **25** | **1,000** |

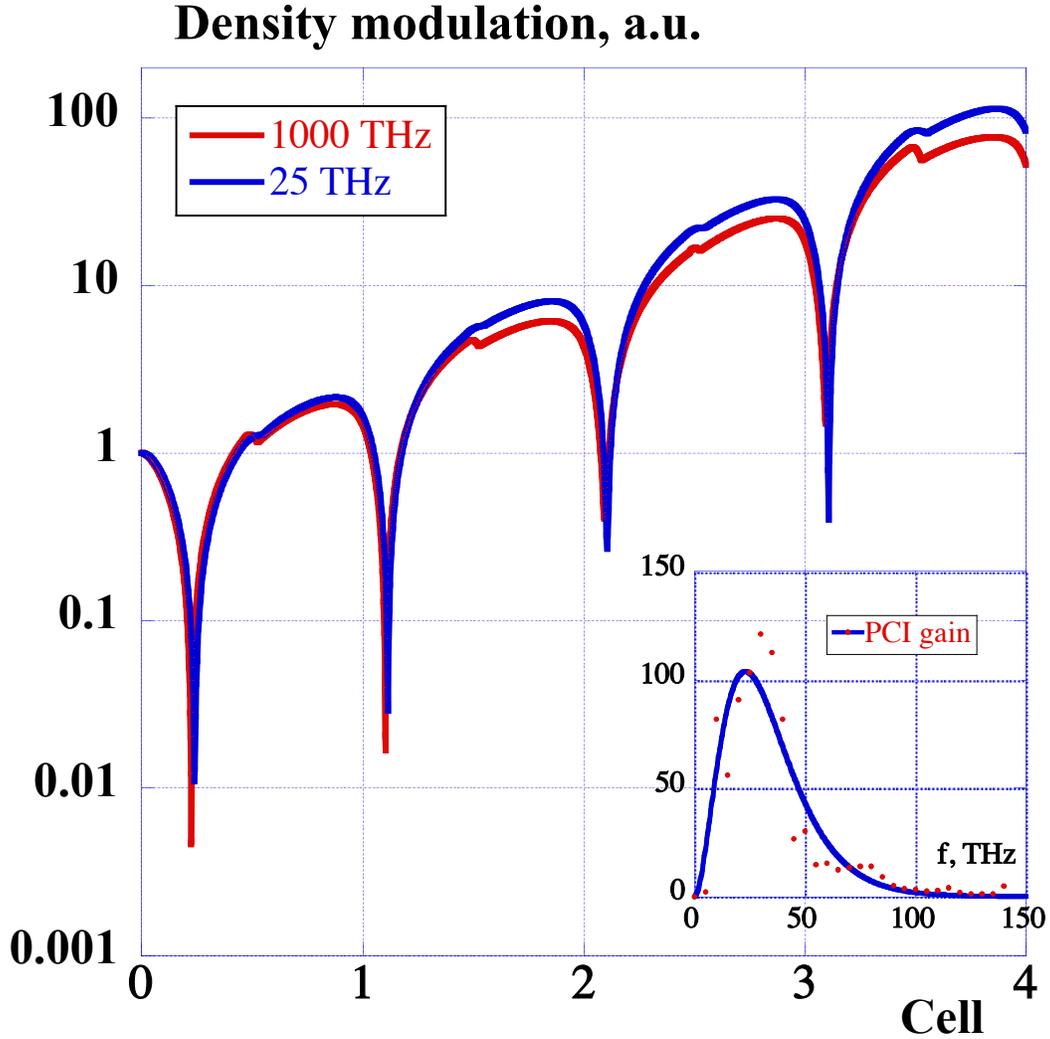

Figure 3. Evolution of density modulation amplitude in 4-cell PCI periodic lattice with parameters shown in Table 1: Blue line, *Case 1*, gain = 114; Red line, *Case 2*, gain=75. The PCI gain spectrum for Case 1 is shown in a clip in the bottom-right corner: red dots are FFT of the amplified shot noise and the blue curve is a smooth fit of the simulated gain.

Each cell consists of a drift section with a length of *2l* between the two focusing solenoids. Initial conditions for this simulation included shot noise and a weak longitudinal density modulation. The beam envelope at the entrance and solenoid strengths are selected to provide the designed beam envelope with waists $a_o$ in the middle of each cell. Amplitude of density modulation, $\tilde{g}_k$, is tracked as a function of the propagation distance. The 3D simulations confirmed our expectations that PCI, as shown in clip in Fig. 3, is a broad-band instability, diminishing both at low and high frequencies.

Fig. 3 clearly demonstrates the nature of this instability: as plasma oscillations progress, the density modulation is transferred into the velocity modulation (at locations of the density minima), and then the velocity modulation generates an increased amplitude of density modulation, but with the opposite sign. This cascade-type growing plasma oscillation is the origin of the name for this instability—PCI.

The growth rates are also in a reasonable agreement. While expected growth in 4-cells (eq. (6) corrected by Landau damping) of 81 (Case 1) and 37 (Case 2) is lower than 114 (Case 1) and 75 (Case 2) predicted by 3D simulations, the difference is exaggerated by the exponential nature of the growth. Indeed, this difference corresponds to 9% and 19% difference in growth rate per cell. The latter is likely the result of using Gaussian energy distribution in simulation instead of κ-1 distribution used in analytical studies. Hence, we concluded that simulations are in reasonable agreement with the analytical estimations.

In contrast to 1D analytical description, the 3D SPACE provides information about spatial shape of the density modulation. Fig. 4 illustrates such features as bending of the modulation wave-fronts caused by the slowing down of electrons with large amplitude of transverse oscillation[7].

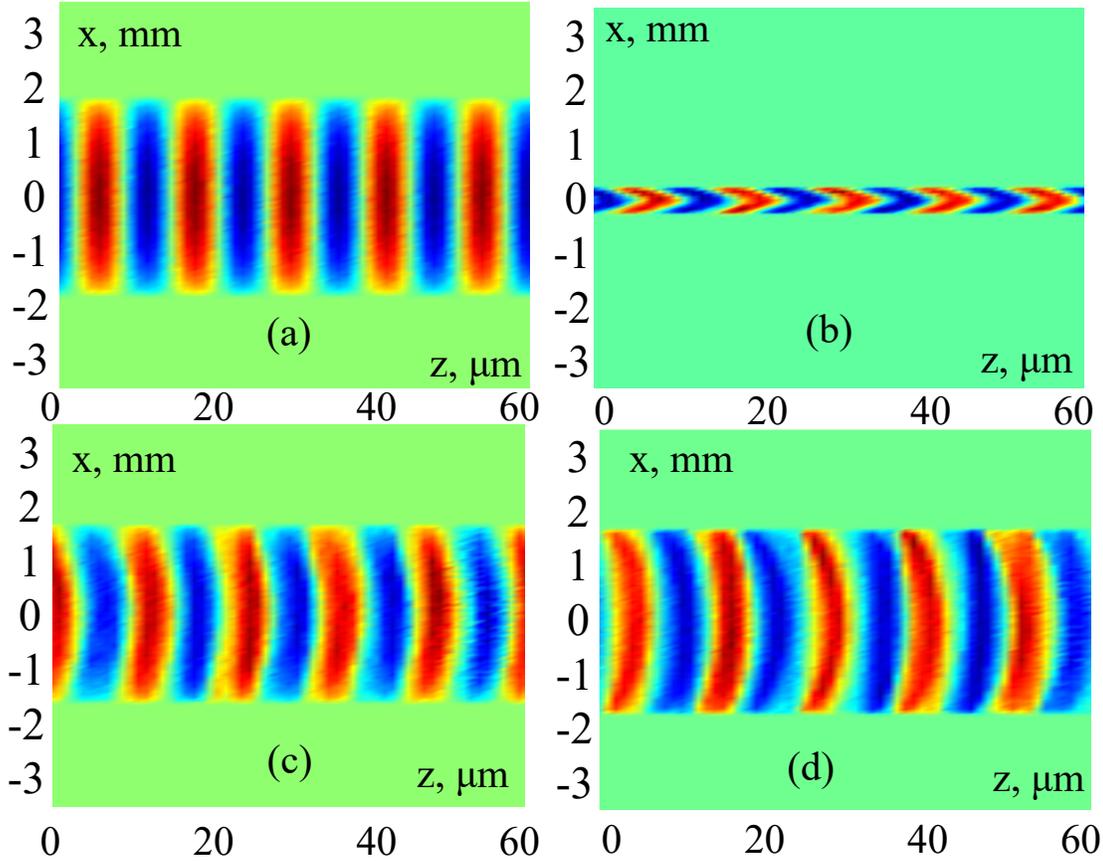

Figure 4. Evolution of the 3D profile of the e-beam density modulation at 25 THz (Case 1). Only AC portion of the density is shown: red color corresponds to increased density and blue to reduced beam density. Color density is normalized to maximum values to accommodate its exponential growth. (a)—at the system entrance; (b)—in the middle of the 2nd cell (beam waist); (c)—after the 3rd cell, and (d)—after 4-cell system.

---

[7] Longitudinal velocity of a particle traveling with angle $\theta$ with respect to the beam direction is reduced as $\mathbf{v}_s = \mathbf{v}/(1+\theta^2/2)$, where $\mathbf{v}$ is the total velocity of the particle.

*Experimental demonstration.* We experimentally observed broad-band PCI at frequencies ~ 0.5 THz and ~ 10 THz using linear superconducting (SRF) accelerator, shown in Fig. 5, built for the CeC experiment [114]. We used 400 psec electron bunches with various charges generated in the 1.25 MV SRF photocathode gun. Two room temperature RF 500 MHz cavities were used to correct the gun's RF curvature and reduce energy spread in the bunch to 0.01% level. Strong-focusing aperiodic lattice comprised of six solenoids in the low energy beam transport (LEBT) and provided beam envelope shown in Fig.5. To observe the density modulation in the beam we transformed our linac and dipole beam-line into a time-resolving system with sub-psec resolution. We operated the linac at zero crossing with low accelerating voltage, $V$ ~100-200 kV, to correlate particle's energy with the arriving time $E = E_o + eV\sin\omega_L t$. The 45º dipole and the profile monitor 4 served as the energy spectrometer. The measured energy distribution was a carbon copy of the bunch's time profile.

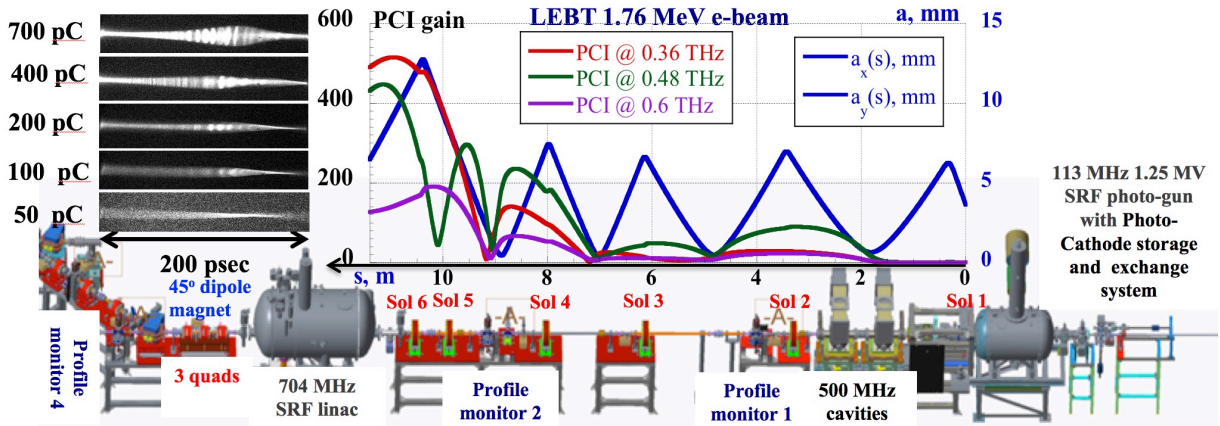

Figure 5. Layout of the CeC accelerator (right to left): the SRF electron gun, two bunching RF cavities, the LEBT line equipped with six solenoids and two profile monitors, followed by the 13.1 MeV SRF linac and a 45º bending magnet beam line (with three quadrupoles and a beam profile monitor). The top graph shows simulated (by code SPACE) evolutions in the LEBT of the beam envelope (*a(s)*, blue line) and PCI gains at frequencies of 0.36 THz (red line), 0.48 THz (violet), and 0.6 THz (green). Simulations were done for 1.75 MeV ($\gamma$=3.443), 0.7 nC, 0.4 nsec electron bunch with 1 μm normalized slice emittance and 0.01% slice RMS energy spread. Clip in the left-top corner shows time-resolved bunch profiles measured by the system for various charges in 400 psec electron bunch.

Time resolution of our measurement system depends on the linac voltage. We tested the system with the voltage from 50 kV to 200 kV and found that the best data quality was obtained at 100 kV setting: at lower voltages we did not have sub-psec time resolution, but at higher voltages diminishing electron bunch density was reducing the signal to noise ratio. Since density modulation grows from the random shot noise, averaging images was not an option—it would erase the modulation generating useless smooth profiles. Hence, it was critical to measure time structure of a single bunch in a single shot. The 100 kV setting provided 0.373 mm/psec scaling at the profile monitor and, correspondingly, 7 pixel/psec at the digital camera. Combined with YAG crystal in the profile monitor spatial resolution ~ 0.1mm, this provided for attainment of sub-psec time resolution—subject for fine focusing using three quadrupoles in front of the dipole magnets.

The 1" (25.4 mm) YAG crystal was intercepting electrons with momentum spread of ±0.93%, corresponding to 65 psec time slice in the bunch center. The 65 psec slice corresponds to ±7° of the RF phase in the 704 MHz SRF linac, which guarantees linearity of the above scaling. Fig. 6 shows a few selected density profiles measured by our system as well as their spectra.

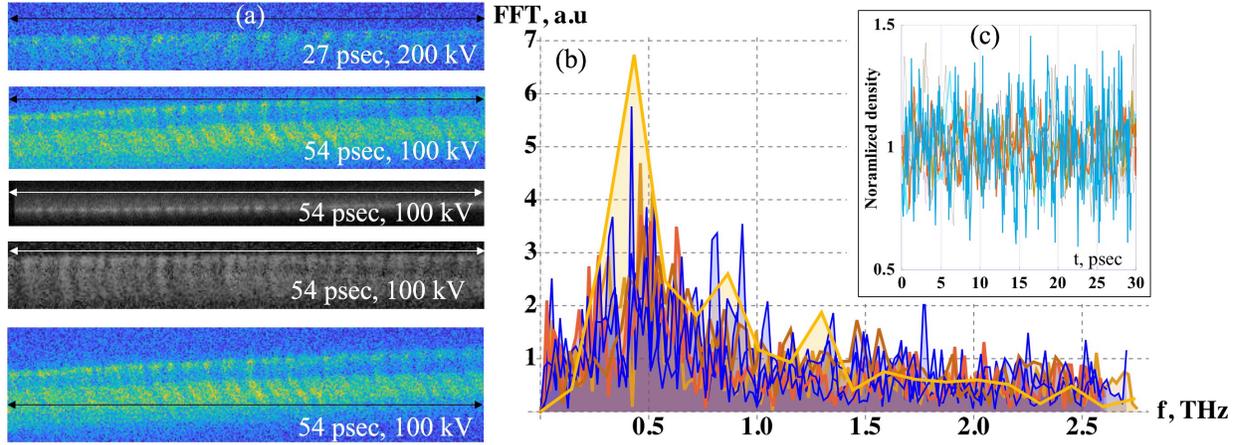

Figure 6. (a) Measured time profiles of 1.75 MeV electron bunches emerging from LEBT. Charge per bunch was from 0.45 nC to 0.7 nC; (b) Seven overlapping spectra of measured bunch density modulation and PCI spectrum simulated by SPACE (slightly elevated yellow line); (c) Clip shows a 30-psec fragment of seven measured relative density modulations.

We observed a very large, ±50%, density modulation that can be seen both in the captured images, or in the density profiles in Fig.6 (a) and (c). Fig. 6 (b) shows measured bunch spectra, which compares very well with the simulation: a broadband PCI gain peaking at ~ 0.4 THz. The high and low frequency noise floor is determined by the noise in the CCD camera and it is ~ 100 higher than natural shot noise in the beam. The calculated correlation length of the density modulation ~ 1.5 and large spectral bandwidth of instability $\Delta f/f$ ~ 1 are in good agreement.

While all our measurements were in good agreement with our simulations, to clarify that the observed structures were indeed caused by the PCI we preformed following tests:

(a) We ran simulations of beam dynamics including and excluding wakefields from the beam-line components in LEBT (calculated using codes CST [115], Echo [116], and ABCI [119]) and found signature at frequencies above 0.01 THz;

(b) We checked that modulation at frequencies ~ 0.5 THz does not exist in the laser pulses driving the beam—again, the laser spectral line was well within 0.05 THz;

(c) We studied dependence of the density modulation on the electron beam charge. As predicted by simulation, there is a very strong dependence of density modulation amplitude on the bunch charge. Instability disappears for charge per bunch below 100 pC—see Fig. 5.

Nominally, for CeC operations, we compress electron bunches 20-fold in LEBT by applying energy chirp by 500 MHz cavities. This provided us with an opportunity to observe PCI at frequencies ~ 10THz. Simulations using Impact-T code [117] clearly indicated broad-band PCI with gain ~ 15-20 at and around frequency of 10 THz. Noise in our CCD camera and absence of any dedicated time-resolving equipment did not allow us to attain resolution ~ 30 fsec necessary to optically observe density modulation at these frequencies. The CeC FEL amplifier [116,118] operating at wavelength of 31 μm (e.g. ~ 10 THz) is equipped with broad-band IR diagnostics. We used this diagnostic to measure radiation power from our 45-degree bending magnet[8] and found that it exceeded level of natural (e.g. originated from Poisson statistics) spontaneous radiation ~300±100 fold, again is in reasonable agreement with the predicted increase in the amplitude of the density modulation ~ 15-20-fold. While not being direct measurements of the bunch structure, this measurement provided additional evidence of PCI and its ability to generate high frequency modulation in electron beams.

In conclusion, we would like to announce the discovery of a novel microbunching instability occurring in charged particle beams propagating along a straight trajectory—Plasma-Cascade Instability. PCI can be both a menace and a blessing—it can strongly intensify longitudinal micro-bunching originating from the beam's shot noise or an external source. Resulting random density and energy microstructures in the beam can become a serious problem for generating high quality electron beams. On the other hand, such instability can drive novel high-power sources of broadband radiation or can be used as a very broad band low-noise amplifier. For example, PCI can serve as a broadband amplifier in the CeC scheme [75] or used for boosting power of THz and PHz radiation [119-124].

Authors would like to thank all our colleagues from BNL contributed to the CeC project, with special acknowledgements going to Toby Miller for developing and commissioning high precision beam profile monitor diagnostics, Thomas Hayes, Geetha Narayan and Freddy Severino for their help with using CeC SRF linac for time resolved studies, Patrick Inacker for providing detailed measurements of the laser bandwidth, Dr. Peter Thieberger for pointing on important 3D aspects of the problem, and Dr. Thomas Roser for encouragement and unrelenting support of this research. First author also would like to thank Prof. Pietro Musumeci (UCLA), who mentioned during our discussion about energy conservation in longitudinal plasma oscillations, that modulation of the transverse beam size can violate this perception. His notion was one of the initial signals that modulation of the transverse beam size can cause an instability.

This research was supported by and NSF grant PHY-1415252, DOE NP office grant DE- FOA-0000632, and by Brookhaven Science Associates, LLC under Contract No. DEAC0298CH10886 with the U.S. Department of Energy.

---

[8] Interpretation of measured FEL power is rather convoluted: it can come either from FEL gain or initial noise in the beam. Measuring power radiated by electron beam in a bending magnet is straightforward.